\documentclass[conference]{IEEEtran}
\IEEEoverridecommandlockouts
\usepackage{cite}
\usepackage{amsmath,amssymb,amsfonts}
\usepackage{mathtools}
\usepackage{algorithm,algpseudocode} 
\usepackage{multicol}
\usepackage{graphicx}
\usepackage{textcomp}
\usepackage{xcolor}
\usepackage{booktabs,multicol,caption}
\usepackage{url}

\DeclareMathOperator*{\argmax}{argmax}

\def\BibTeX{{\rm B\kern-.05em{\sc i\kern-.025em b}\kern-.08em
    T\kern-.1667em\lower.7ex\hbox{E}\kern-.125emX}}
\begin{document}


\title{Maximizing Clearance Rate of Reputation-aware Auctions in Mobile Crowdsensing}

\author{Maggie E. Gendy\textsuperscript{1}, Ahmad Al-Kabbany\textsuperscript{1,2,3}, and Ehab F. Badran\textsuperscript{1} \\
	\textsuperscript{1} Department of Electronics and Communications Engineering,\\
	 Arab Academy for Science, Technology, and Maritime Transport, Alexandria, Egypt \\ 
	 \textsuperscript{2}Intelligent Systems Lab, Arab Academy for Science, Technology, and Maritime Transport, Alexandria, Egypt \\ 
	\textsuperscript{3} Department of Research and Development, VRapeutic, Alexandria, Egypt
}


\maketitle

\begin{abstract}
	Auctions have been employed as an effective framework for the management and the assignment of tasks in mobile crowdsensing (MCS). In auctions terminology, the \emph{clearance rate} (CR) refers to the percentage of items that are sold over the duration of the auction. This research is concerned with maximizing the CR of reputation-aware (RA) auctions in centralized, participatory MCS systems. Recent techniques in the literature had focused on several challenges including untruthful bidding and malicious information that might be sent by the participants. Less attention has been given, though, to the number of completed tasks in such systems, even though it has a tangible impact on the satisfaction of service demanders. Towards the goal of maximizing CR in MCS systems, we propose two new formulations for the bidding procedure that is a part of the task allocation strategy. Simulations were carried out to evaluate the proposed methods and their impact on the user utility, under varying number of auctions, tasks, and participants. We demonstrate the effectiveness of the suggested methods through consistent and considerable increases (three times increase, in some cases) in the CR compared to the state-of-the-art.
\end{abstract}


\section{Introduction}
\label{sec:intro}

\begin{figure}[t]
	\begin{center}
		\setlength{\abovecaptionskip}{-5pt plus 0pt minus 0pt}
		\setlength{\belowcaptionskip}{-15pt plus 0pt minus 0pt}
		\includegraphics[width=3.4in]{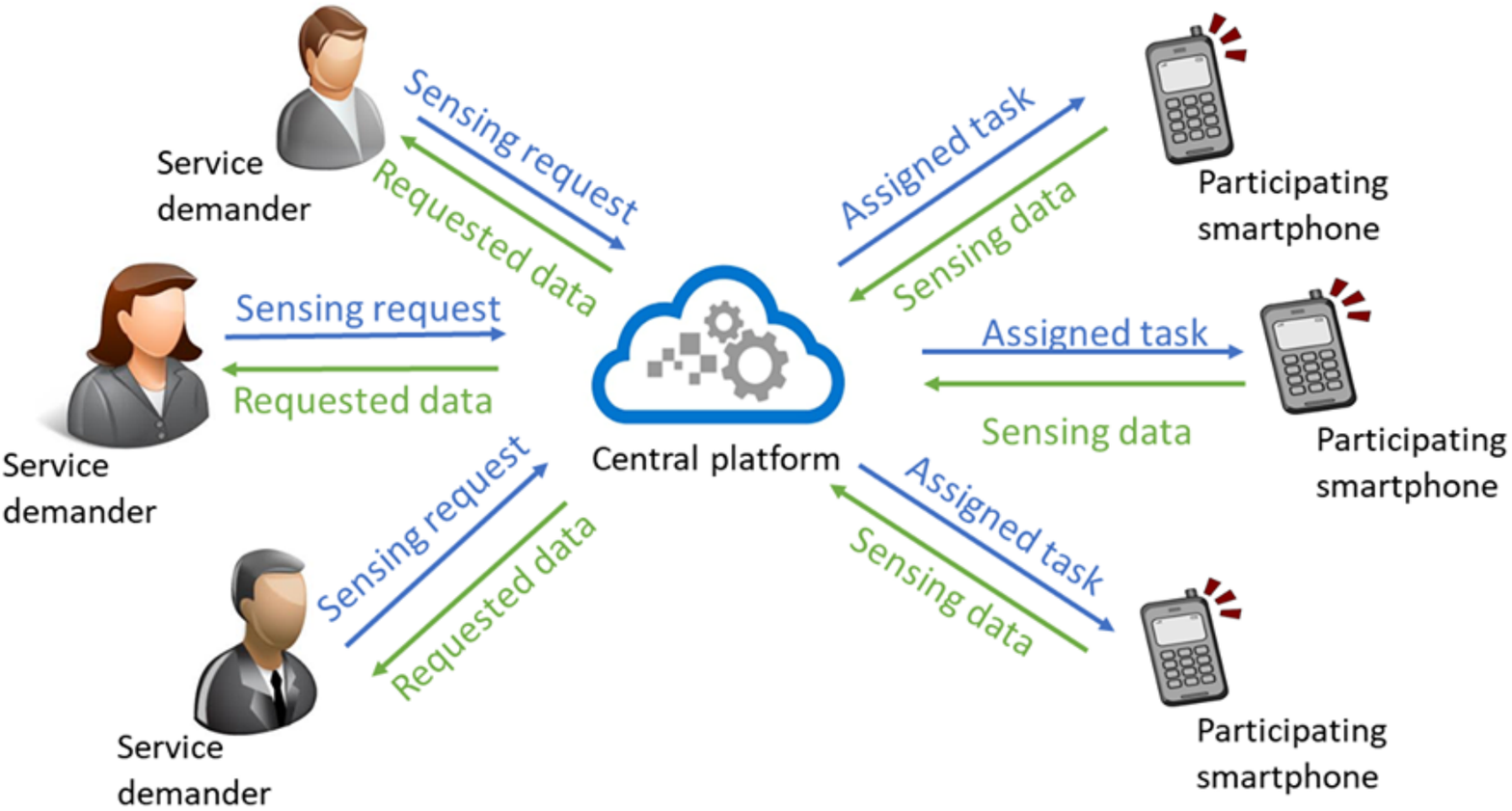}
		\caption{The general architecture of a MCS system.}
		\label{fig:Architecture}
	\end{center}
\end{figure}

The Internet of Things (IoT) denotes a network of possibly
objects, sensors, and devices that are connected through
communications and information infrastructure to provide a
wide variety of services \cite{b1}. With the high-paced advances
in emerging wireless technologies (e.g. 4G/5G) and micro-embedded
highly accurate sensors (such as gyroscope, barometer,
etc.), smartphones and other intelligent mobile devices
do not only act as communication devices, but they can
also collect information and be programmed to transmit such
measurements wirelessly over the internet. With such a pervasive
computing paradigm, it had been stimulating for another
paradigm, the Mobile Crowd Sensing (MCS), to get realized.
The latter is inspired by crowdsourcing, as a distributed
problem-solving model, where a large number of volunteers
get involved in order to solve a complex problem. Through
sensor-equipped mobile devices, MCS is used to acquire local
knowledge and to measure and map phenomena of common
interest \cite{b10,b11}. This has been employed in environmental
applications \cite{b13} (e.g., measuring air quality and noise
levels), in infrastructure research \cite{b15} (e.g. measuring traffic
congestion and road conditions), and in social applications \cite{b23}
(e.g. share restaurant information, and crowding counting).
The general architecture of a MCS system is shown in Fig~\ref{fig:Architecture}.

Centralized MCS applications, which consist of a central
platform and a number of smartphones, involve different forms of sensing, namely, participatory and opportunistic sensing. The former requires
active involvement of users in sensing and decision-making, while the latter automatically detects a state of interest and changes the device state to satisfy an application request. As participatory sensing can support diverse applications, it is widely applied in MCS systems \cite{b25}. 

Throughout the past decade, many crowdsensing platforms
have been proposed such as MOSDEN \cite{b28}, OpenSignal, and LifeMap \cite{b29}. Also, Google has developed a MCS application called Science Journal. These platforms bear both, high potential and limitations of MCS systems \cite{b26, b32}. One of the most critical limitations is the restrained computational and power resources of smartphones. Hence, most smartphone users need a kind of compensation in order to participate. High-paced research has focused on developing effective incentive mechanisms in order to ensure users willingness to share their sensing data. 

In research, incentives can be classified using diverse approaches;
a widely used approach is to classify them as
either monetary or non-monetary \cite{b44}. In relation to the scope
of this work, we limit the discussion to monetary mechanisms,
where the payments to participants can either be static or
dynamic. In the static approach, the amount of reward remains
constant through the whole experiment. The authors of \cite{b43} defined two incentive mechanisms: 1) The
platform-centric model, where the static payment for each
winner is determined by the platform. The process is modeled
as a Stackelberg game--the participants against the platform.
2) The user-centric model, that is a reverse auction approach,
where the platform has no control over payments to each
winner. Also, it was shown that user-centric local search-
based auction is vulnerable to untruthful bidding. This
problem was addressed in the same study by the \emph{Msensing}
auction. Nevertheless, \emph{Msensing} has also shown vulnerability
to users aiming to send malicious information. To address
this challenge, the authors of \cite{b48} proposed a reputation-aware (RA) incentive mechanism (\emph{TSCM}), which is considered to be the RA version of the \emph{Msensing}.

\begin{figure}[t]
	\begin{center}
		\setlength{\abovecaptionskip}{3pt plus 0pt minus 0pt}
		\setlength{\belowcaptionskip}{-15pt plus 0pt minus 0pt}
		\includegraphics[width=3.4in,height=2.3in]{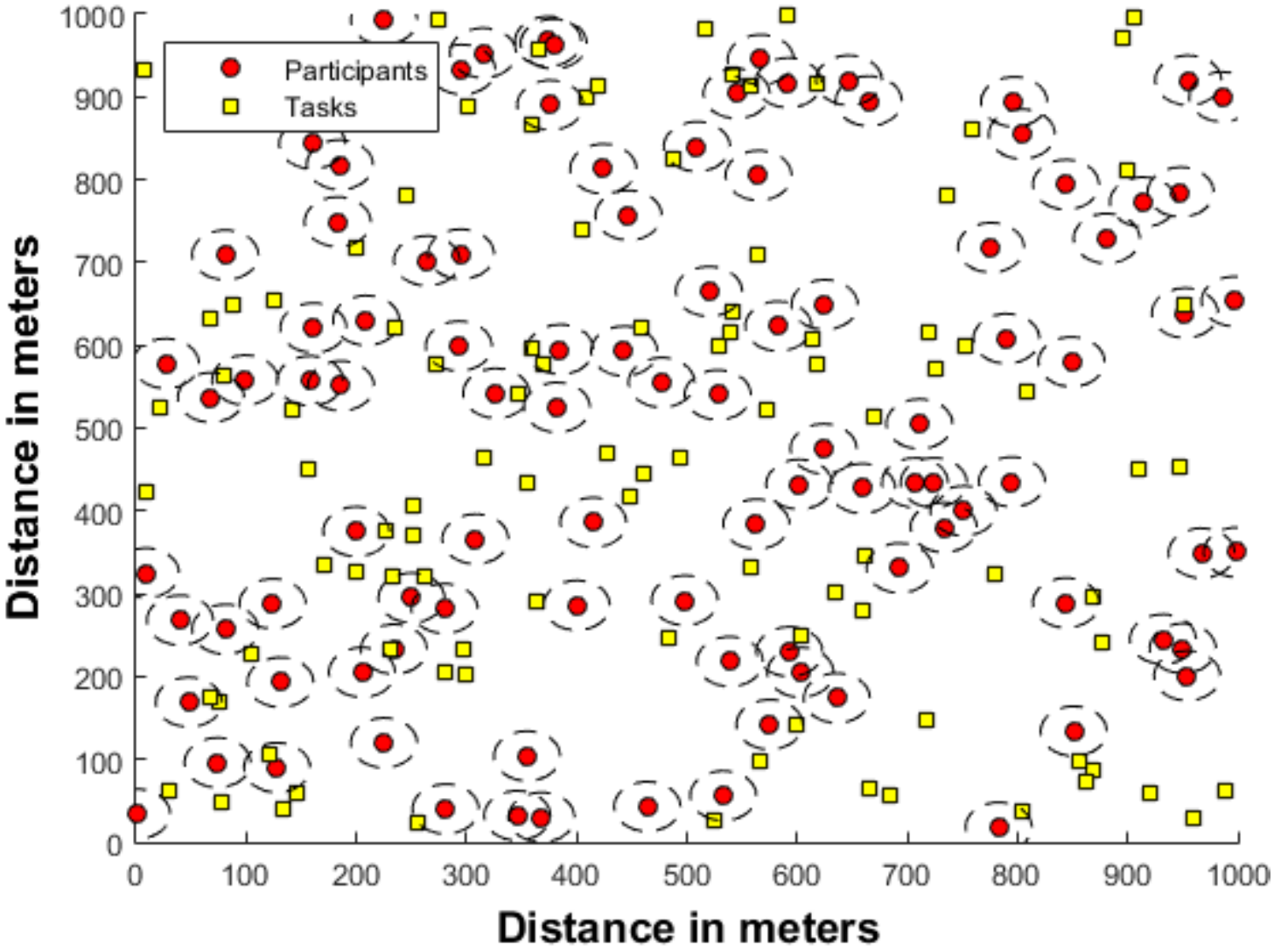}
		\caption{An illustration inspired by \cite{b43} depicting participants (red dots) interested in tasks (yellow squares) within their areas of interest (dashed circles).}
		\label{fig:ParticipantTask}
	\end{center}
\end{figure}

While a considerable body of research has addressed the improvement
of task assignment and user compensation mechanisms as two fundamental stages in MCS system design, much less attention has been given to maximizing the CR--the number of accomplished
tasks--in auction-based campaigns. This is despite the fact
that high CR implies higher system efficiency and quality of service, and maximizes the satisfaction of service demanders. However, the platform usually achieves a low CR inevitably due to many reasons, among which, is the presence of un-bided tasks and/or out-of-reach tasks as illustrated in Fig.~\ref{fig:ParticipantTask}. The limited platform budget is also one main reason for lack of task coverage. In the rest of this paper, we use the terms \emph{clearance rate}, \emph{task completion ratio}, and \emph{task coverage ratio} interchangeably. 

This research addresses the challenge of maximizing the task completion ratio in centralized, participatory MCS systems. The contributions of this paper are summarized as follows:
\begin{enumerate}
\item We propose two new bidding-based task allocation procedures/
strategies for maximizing the platform utility by
maximizing the number of covered tasks in a campaign.
\item We demonstrate a remarkable enhancement in task completion ratios compared to other methods in the recent literature. Also, we present the algorithm analysis and simulations under varying scenarios and conditions. Finally, we identify the drawbacks of the proposed techniques.
\item Although we are aware of \cite{b12}, to the best of our knowledge,
this research is the first to address the maximization
of CR of auction-based participatory sensing systems by proposing novel bidding procedures.
\end{enumerate}

The rest of this paper is organized as follows. A description of the proposed methods is presented in sec.~\ref{sec:proposed}. Results are discussed in sec.~\ref{sec:results}. Finally, conclusions and future work are given in sec.~\ref{sec:conclusion}.

\section{Auctions Based on Descriptive Bidding}
\label{sec:proposed}
In this section, we discuss the proposed \emph{two-stage bid} (2SB) and \emph{per-task bid} (PTB) algorithms. A summary of the symbols and the notations that are used throughout this document is given in Table~\ref{tab1}. Both of the proposed bidding procedures allocate tasks to participants while aiming to maximize the number of the accomplished sensing tasks in a campaign. These greedy algorithms are approximations of the NP-hard problems of task allocation and auction winner selection. For every sensing campaign \cite{b59}:
\begin{itemize}
\item Each smartphone $i \in {1,\cdots,N}$ represents a participant in the auction.
\item The platform sends the details of the $M$ campaign tasks, where tasks are indexed by $j \in {1,\cdots,M}$.
\item All of the participants should take part in the bidding process for the tasks they are interested in, and each bidder should at least bid on one task.
\item Winner selection and payment determination algorithms are then used to find the sets $S$ and $\{P\}$.
\end{itemize}
Unlike previous techniques that do not take budget constraints and/or the CR into consideration \cite{b48}, and/or assume a constant-yet-arbitrary budget \cite{b12}, our proposed bidding procedure link the number of campaign tasks to the platform budget, and maximize the task completion ratio. 

\begin{table*}[t]
	\centering
	\begin{center}
		\caption{A summary of symobls and notations used throughout the document}
		\label{tab1}
	\begin{tabular}{p{1cm}|p{7cm}||p{1cm}|p{6.5cm}}
		\toprule
		\textbf{Symbol} & \textbf{Meaning} & \textbf{Symbol} & \textbf{Meaning} \\ \hline
		$M$, $N$ & Number of tasks, Number of participants & $\mathcal{B}$  & Platform budget \\ \hline
		\addlinespace[0.25em]
		$P$ & Set of participants & $\mathcal{V}$  & Sum of campaign tasks values \\ \hline
		\addlinespace[0.25em]
		$V_j$ & Value of task $j$ & $\{V\}$ & Values of set of tasks \\ 
		\hline
		\addlinespace[0.25em]
		 $T$ & Set of all campaign tasks & $T_i$  & Set of tasks done by user $i$ \\ \hline
		\addlinespace[0.25em]
		 $M_i$ & The number of tasks done by user $i$ & $R_i$  & Reputation of user $i$ \\
		\hline
		\addlinespace[0.25em]
		$R$ & Participants reputation set &  $B^c_i$ & Collective bid for user i \\ \hline
		\addlinespace[0.25em]
		$B_{ij}$  & bid of user i for task j (descriptive bid) &  $S$ & Set of primary winners (selected users) \\ \hline
		\addlinespace[0.25em]
		$\{B^c\}$ & Set of collective bids by participants & $S^{s}$  & Set of secondary winners (selected users) \\ \hline
		\addlinespace[0.25em]
		$\mathcal{P}$  & Sum of all payments to primary winners &  $P_i$ & Payment to winner $i$ \\ \hline
		\addlinespace[0.25em]
		$T^s_i(S^s)$ & Set of tasks allocated for secondary winner $i$ over the set $S^s$ & $\{P\}$ & The set of payments
		\\ \hline
		\addlinespace[0.25em]
		 $X_{ij}$ & Flag if the user $i$ is interested in task $j$ or not, $X_{ij}\in\{0,1\}$ & $V^R_i(S)$  & Reputational value for user $i$ over set $S$ \\ 
		\bottomrule
	\end{tabular}
    \end{center}
\end{table*}

\vspace{10pt}
\noindent {\emph{\textbf{Two-stage Bidding.}}\hspace{5pt}
For a campaign with a set of tasks $T$, with cardinality $|T|=M$, every potential participant, who is interested in a subset of tasks $T_i \subset T$, sends two types of bids to the platform, namely, a \emph{collective bid} and a \emph{descriptive bid}. The former is the classical form of bidding, commonly discussed in the literature, that resembles a wholesale or bidding in bulk, where the user asks for one collective payment in return for all the tasks in $T_i$. In descriptive bidding, however, a participant sends a list of tasks and a separate bid for each of them. Throughout this document, we refer to this list as \emph{the list of per-task user bids}. The summation of the per-task user bids for the user $i$ is given by:
\begin{equation}
B_i = \sum_{j=1}^{M_i}X_{ij} . B_{ij}. 
\label{eqn:sumbids}
\end{equation}
Unless the bidder is interested in only one task, the sum of the descriptive bids is usually more than the collective bid \cite{b36}. 

The steps of the reputation-aware 2SB is given by algorithm listing~\ref{algo:rep2SB} below. Following previous work in the literature \cite{b48},  the algorithm starts by calculating the marginal contribution (or marginal value) for each participant, as formulated by \cite{b48}, and then subtracts their collective bids from the resultant value (line 3). Afterwards, tasks are allocated to a set of winners, called primary winners (lines 5,6), that are chosen such that the budget (same as platform utility in \cite{b48}) is maximized. The mathematical expression of the budget is given in \ref{eqn:bpv}. It is worth mentioning that for reputation-unaware (RU) bidding, $R_i=1$ for user $i$.

Following the payment calculation for the primary winners, and given the budget of the platform, we can determine the remaining budget that is available--before getting a negative utility--to accomplish the tasks that had not been covered by primary winners. This is given by:
\begin{equation}
\mathcal{B} = \mathcal{V}-\mathcal{P}.
\label{eqn:bpv}
\end{equation}

Unless the set of $M$ tasks have been covered by the primary winners, the platform proceeds to the second stage of the algorithm. Using the descriptive bids, the platform determines another set of winners, called the secondary winners, to whom the uncovered tasks are allocated  (lines 33-37). On the expense of the budget, the platform pays the secondary winners according to their descriptive bids in order to achieve a higher CR. This happens because either they are the only bidders for some tasks or because of their unique location near to particular tasks afar from the crowd. Moreover, unlike primary winners, while assigning uncovered tasks to secondary winners, the platform ensures that a task would not be covered more than once, for better budget management (lines 26-32 for budget computation, lines 38-43 for budget update).

\begin{algorithm*}
	\caption{Reputation-aware two-stage bid}
	\label{algo:rep2SB}
	\begin{multicols}{2}
		\begin{algorithmic}[1]
			\Procedure{Get Primary Winners}{$\{V\}$, $\{B_p\}$}
    				\State {$S \leftarrow \Phi$}, {$S_{tasks} \leftarrow \Phi$}
				\vspace{0.25em}
    				\State $h \leftarrow \argmax\limits_{p \in P}{V_p^R(S)-\frac{b_p}{R_p}}$
				\vspace{0.25em}
				 \While {$\frac{b_h}{R_h} < V_h^R \hspace{2pt} \bigwedge \hspace{2pt} S \neq P$}
					\vspace{0.25em}
				 	\State $S \leftarrow S \cup {h}$ \hspace{2pt},\hspace{2pt} $S_{tasks} \leftarrow T_h$
					\vspace{0.25em}
					\State $h \leftarrow \argmax\limits_{p \in P \setminus S}{V_p^R(S)-\frac{b_p}{R_p}}$
					\vspace{0.25em}
				\EndWhile
			\EndProcedure
			\vspace{0.25em}
			\Procedure{Get Winners Payments}{$\{V\}$, $\{B_p\}$, $S$}
				\For{$p \in P$}
					\vspace{0.25em}
        					\State $P_p \leftarrow 0$
					\vspace{0.25em}
      				\EndFor
				\For{$i \in S$}
					\vspace{0.25em}
        					\State $S' \leftarrow S \setminus\{i\}$,$\Theta = \Phi$
					\vspace{0.25em}
					\Repeat
						\State $q \leftarrow \argmax\limits_{\mathcal{V} \in S' \setminus \Theta} (V_\mathcal{V}^R(\Theta) - \frac{b_\mathcal{V}}{R_\mathcal{V}})$
						\vspace{0.25em}
						\State $P_i \leftarrow \max(P_i,\min\{V_i^R(\Theta)-(V_q^R(\Theta)-\frac{b_q}{R_q})\})$     				\vspace{0.0025em}
						\State $\Theta \leftarrow \Theta \cup \{q\}$
						\vspace{0.25em} 
					\Until{$\frac{b_q}{R_q} \geq V_q^R \bigvee \Theta = S'$}
					\vspace{0.25em}
      				\EndFor
				\vspace{0.25em}
			\EndProcedure
			\vspace{0.25em}
			\State $\mathcal{B} = \mathcal{V} - \mathcal{P}$
			\vspace{0.25em}
			\If{$S_{tasks} \neq T$}
				\State $S^{s} = \Phi$
				\State $T^s = \Phi$
				\For{$p \in P \setminus S$} 
					\vspace{0.25em}
					\For{$t \in T_p$}
						\If{$t \in S_{tasks}$}
							\State $T_p = T_p \setminus t$
						\ElsIf{$t \in T \setminus S_{tasks}$}
							\State $P^{s} \leftarrow P^{s} \cup \{p\}$
						\EndIf
					\EndFor	
				\EndFor
			
				\vspace{0.25em}
				\State $h \leftarrow \argmax\limits_{p \in P^{s}} (V_p^R(S^{s})-\frac{B_p(S^{s})}{R_p})$
				\vspace{0.25em}
				\While {$\frac{B_h}{R_h} + (R_h \times \mathcal{B}) \geq 0\hspace{2pt}\bigwedge \hspace{2pt} S^{s} \neq P^{s}\hspace{2pt} \bigwedge \hspace{2pt} S_{tasks}\neq T$}
					\vspace{0.25em}
					\vspace{0.25em}
					\State $S_{tasks} \leftarrow S_{tasks} \cup T^s_h(S^s)$
					\vspace{0.25em}
					\State $S^{s} \leftarrow S^{s} \cup \{h\}$
					\vspace{0.25em}
				 	\For{$p \in P^{s} \setminus S^{s}$} 
						\vspace{0.25em}
						\For{$t \in T_p$} 
						\vspace{0.25em}
							\If{$t \in T^s_h$}
							\State $T_p = T_p \setminus t$
							\EndIf
						\EndFor
					\EndFor
					\vspace{0.25em}
					\State $\mathcal{B} = (\mathcal{B} \times R_h)-\frac{B_h}{R_h}$
					\State $h \leftarrow \argmax\limits_{p \in P^{s} \setminus S^{s}} (V_p^R(S^{s})-\frac{B_p(S^{s})}{R_p})$
				\EndWhile
			\EndIf
			\State $Outlier\_Detection(S,S^{s})$
			\For{ $s \in \{S \cup S^{s}\}$} 
				\State update $R_s$
			\EndFor
			\State \Return ($S,S^{s},\mathcal{P}, R$)
		\end{algorithmic}
	\end{multicols}
\end{algorithm*}

\vspace{10pt}
\noindent {\emph{\textbf{Per-task Bidding.}}\hspace{5pt}
The 2SB algorithm starts off the auction with collective bidding and then handles the uncovered tasks using descriptive bidding.  The per-task bidding (PTB) procedure, however, manages the whole auction, from the beginning, by descriptive bids. Consequently, this procedure does not require the user's collective bid.

\section{Results and Discussion}
\label{sec:results}
The simulation is done in an area of ($1000$ $m$ $\times$— $1000$ $m$) in which participants and tasks are are uniformly distributed. Each participant is surrounded by an area of interest of $30m$ radius as depicted in Fig.~\ref{fig:ParticipantTask}. Following \cite{b43,b48}, the value of each task and the participants' collective bids vary uniformly in [1,5] and [1,10] respectively. Similarly, the per-task bids vary uniformly in the range [$V_j - \alpha$, $V_j + \alpha$], and we set $\alpha =2$ in our simulations. The participants' reputations are varied uniformly from $0.6$ to $0.9$. To evaluate the effectiveness of our algorithms, we compare the performance of the reputation-aware and reputation-unaware versions of the 2SB and the PTB algorithms to two algorithms from the literature, namely, \emph{Msensing} \cite{b43} and \emph{TSCM} \cite{b48} as representatives of reputation-unaware and reputation-aware techniques respectively. We use two metrics in our evaluation, the tasks completion percentage and the user utility. Three factors are considered in our simulations which are: the number of auctions, the number of tasks, and the number of participants. Table~\ref{tab2} summarizes simulated scenarios and their corresponding parameter values.

\subsection{Simulation Results for Two-stage Bids}
\subsubsection{The impact of varying the number of auctions on the CR}
First, we investigate the impact of the number of held auctions on the performance of the platform. As summarized in Table~\ref{tab3}, for both reputation-aware and reputation-unaware 2SB, the CR achieved by the proposed method is close to three times higher than \emph{TSCM} and \emph{Msensing}. The average percentage of tasks completion is nearly constant, regardless the number of auctions. 

\begin{table*}[t]
\begin{center}
\caption{A summary of the different simulated scenarios and their corresponding parameter values.}
\setlength{\abovecaptionskip}{5pt plus 0pt minus 0pt}
\setlength{\belowcaptionskip}{-10pt plus 0pt minus 0pt}
\includegraphics[width=6in]{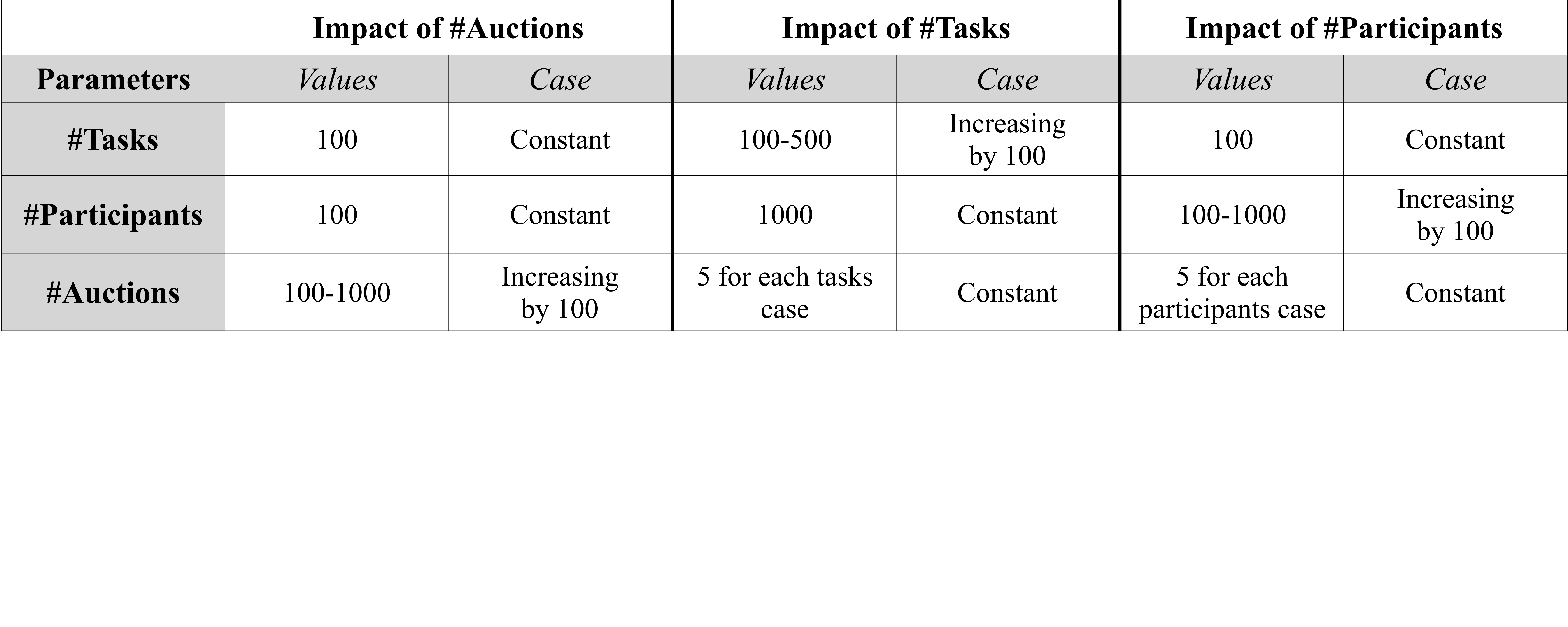}
\label{tab2}
\end{center}
\end{table*}

\begin{table}[t]
	\centering
	\begin{center}
		\caption{Average task completion percentages of 2SB-RA, 2SB-RU, PTB-RA, and PTB-RU compared to \emph{TSCM} and \emph{Msensing}.}
		\label{tab3}
	\begin{tabular}{p{1.6cm}||p{0.75cm}|p{0.75cm}|p{0.75cm}|p{0.75cm}|p{0.75cm}|p{0.75cm}}
		\toprule
		\textbf{Technique} & \emph{2SB-RA} & \emph{PTB-RA} & \emph{TSCM} & \emph{2SB-RU} & \emph{PTB-RU} & \emph{Msen-sing}\\ \hline
		\addlinespace[0.25em]
		\textbf{\%Completion} & \emph{14\%} & \emph{11\%} & \emph{5\%} & \emph{24\%} & \emph{24\%} & \emph{7\%}\\
		\bottomrule
	\end{tabular}
    \end{center}
\end{table}



\subsubsection{The impact of varying the number of tasks on the CR}
Fig.~\ref{fig:ParamNumTask2SB} shows that the proposed method, with its reputation-aware and reputation-unaware versions, consistently achieves higher CRs than \emph{TSCM} and \emph{Msensing}. When the number of available tasks ($M$) increases, the CR is expected to increase in all of the algorithms. Meanwhile, the proposed algorithm maintained the highest CR among the techniques under consideration. This is because other techniques aim at maximizing the user and the platform utility through only one stage of bidding (collective bidding), while our algorithm proceeds to another round of bidding to make the best out of the platform budget and better satisfy service demanders. This is done without compromising the user utility as will be shown below. 



\begin{figure}[t]
\begin{center}
\setlength{\abovecaptionskip}{5pt plus 0pt minus 0pt}
\setlength{\belowcaptionskip}{-10pt plus 0pt minus 0pt}
\includegraphics[width=3.4in]{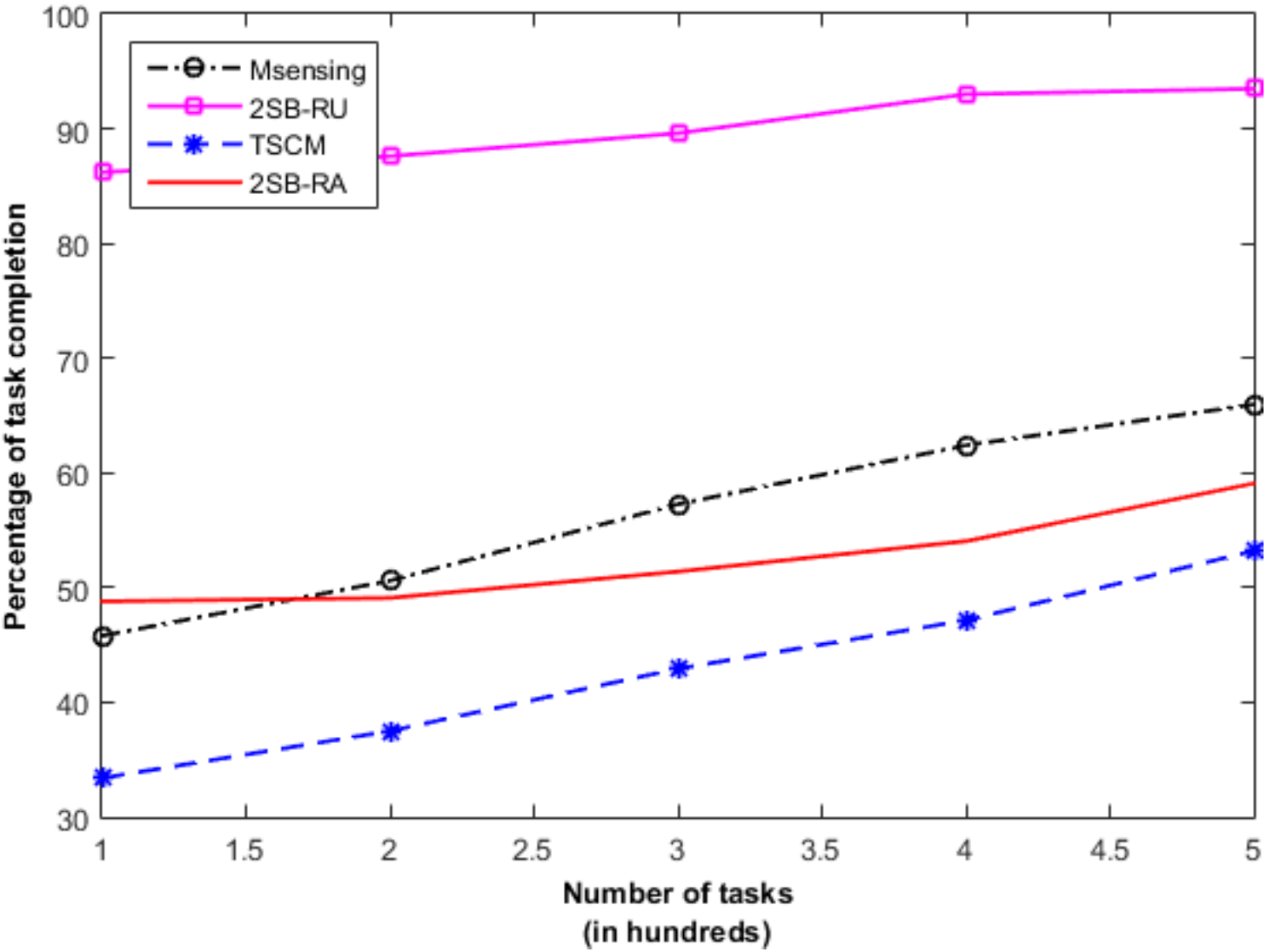}
\caption{The impact of varying the number of tasks on the performance of 2SB with its reputation-aware (RA) and reputation-unaware (RU) versions.}
\label{fig:ParamNumTask2SB}
\end{center}
\end{figure}

\subsubsection{The impact of varying the number of participants on the CR}
As shown in Fig.~\ref{fig:ParamNumPart2SB100}, when the number of participants increases, more candidates compete to be chosen by the platform. Hence, the probability of finding a set of candidates with high marginal contribution, within the platform budget, increases. Thus, the CR increases. Our proposed methods attain consistently higher CR, though, compared to the other techniques. 


\begin{figure}[t]
	\begin{center}
		\setlength{\abovecaptionskip}{5pt plus 0pt minus 0pt}
		\setlength{\belowcaptionskip}{-10pt plus 0pt minus 0pt}
		\includegraphics[width=3.4in]{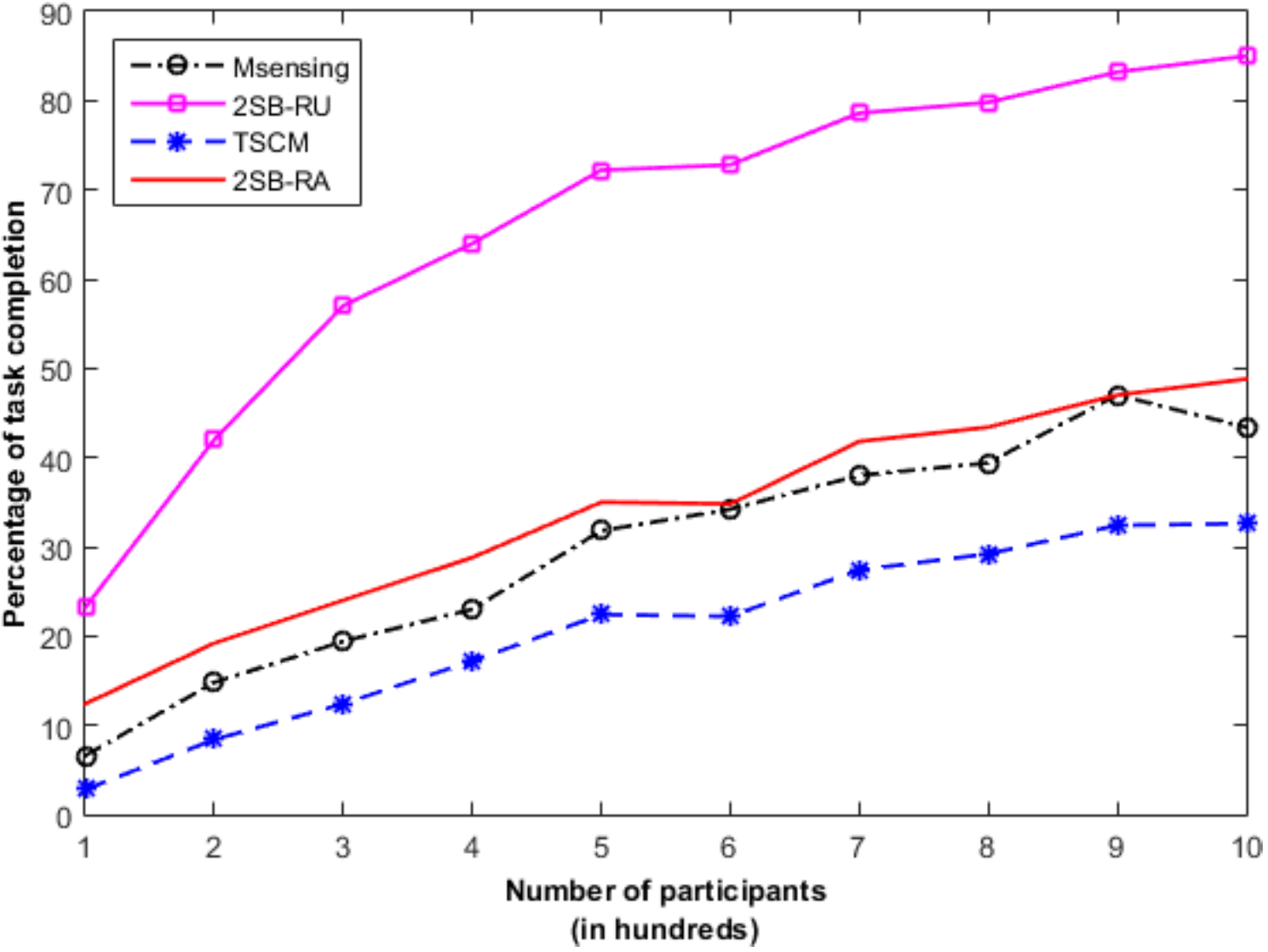}
		\caption{The impact of varying the number of participants on the performance of 2SB.}
		\label{fig:ParamNumPart2SB100}
	\end{center}
\end{figure}

\subsubsection{Concerning user utility}
The proposed bidding procedure implies two types of user utility: The primary users utility which is the difference between the payment and the cost of each winner, and is given as
\begin{equation}
U_{i} = P_i - C_i  \hspace{5pt};\hspace{5pt} i \in S,
\label{eqn:Ui}
\end{equation}
and the secondary users utility which is given by: 
\begin{equation}
\widetilde{U_j} = \left\{ 
  \begin{array}{l l}
    B_j - C_j & \quad \text{if  $B_j > C_j \hspace{5pt} \bigwedge \hspace{5pt} j \in S^{s}$},\\
    0 & \quad \text{otherwise.}
  \end{array} \right.
  \label{eqn:piecwise}
\end{equation}
We formulate the overall utility as their sum, which is given by:
\begin{equation}
U_{ov} = \sum_{i \in S} U_i + \sum_{j \in S^{s}}\widetilde{U_j}.
\label{eqn:Uoverall}
\end{equation}
Concerning the cost for each secondary winner:
\begin{itemize}
\item If the descriptive bid is higher than or equal the collective bid, the winner's cost is equal to the collective bid.
\item Otherwise, the winner's cost is equal to the descriptive bid, and the user utility is equal to zero.
\end{itemize}
Hence, our two-stage bid algorithm is individually rational \cite{b43}, meaning that each user can never have a negative user utility. 
Fig.~\ref{fig:AvgUserUtility} depicts the average user utility which is given as
\begin{equation}
U_{avg-{user}} =\frac{\sum_{i \in S} U_i}{|S|} + \frac{\sum_{ j \in S^{s}} \widetilde{U_j}}{|S^{s}|}\cdot
\label{eqn:FracSigmas}
\end{equation}


\begin{figure}[t]
	\begin{center}
		\setlength{\belowcaptionskip}{-15pt plus 0pt minus 0pt}
		\includegraphics[width=3.4in]{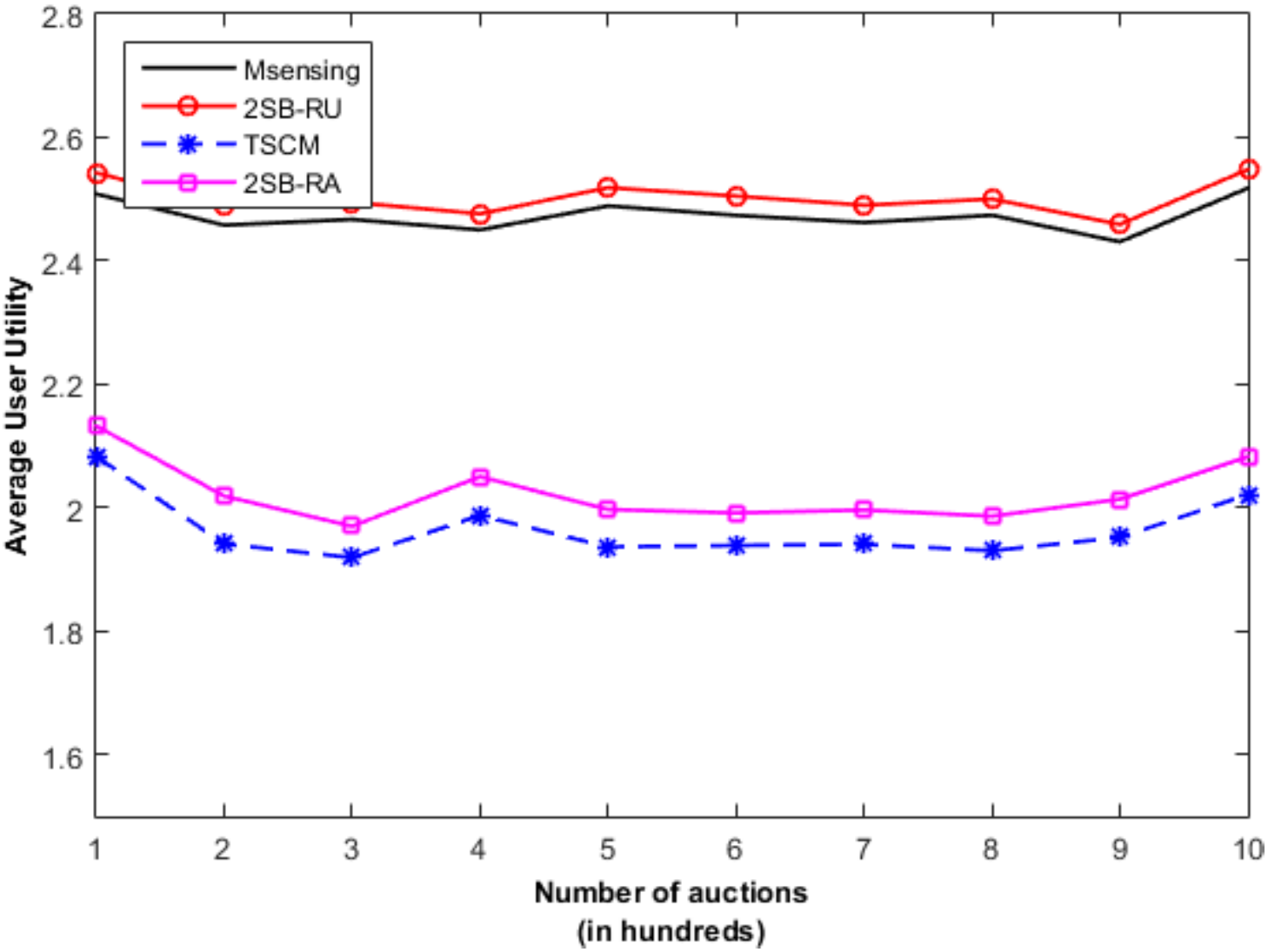}
		\caption{Comparing the average user utility attained by the 2SB to that attained by recent algorithms in the literature.}
		\label{fig:AvgUserUtility}
	\end{center}
\end{figure}

\subsection{Simulation Results for Per-task Bidding}
In the following simulations, we used the same parameters, and investigated the impact of the same factors as in the 2SB algorithm, which are summarized in Table~\ref{tab2}. With regards to the impact of changing the number of auctions on the CR, similar to the 2SB case, the average percentage of tasks completion has been found to be nearly constant regardless the number of held auctions, and is given in Table~\ref{tab3}.


\subsubsection{The impact of varying the number of tasks on the CR}
As we can see in Fig.~\ref{fig:ParamNumTaskPTB}: as the number of tasks increases, the performance of the PTB algorithm deteriorates. This is because the summation of the descriptive bids list (for one participant) increases, leading to a negative marginal contribution; thus the user is excluded by the platform.

\subsubsection{The impact of varying the number of participants on the CR}
As mentioned earlier, increasing the number of participants generally leads to increasing the CR, since the platform has a richer pool of choices. Meanwhile, since the PTB uses descriptive bids only, which is budget-demanding, the 2SB and \emph{TSCM} attain higher completion ratios on average. When the impact of the budget constraints is more than the impact of the increasing number of participants, we find a CR that rises with a very small rate as depicted in Fig.~\ref{fig:ParamNumPartPTB}.


\begin{figure}[t]
	\begin{center}
		\setlength{\abovecaptionskip}{5pt plus 0pt minus 0pt}
		\setlength{\belowcaptionskip}{-10pt plus 0pt minus 0pt}
		\includegraphics[width=3.4in]{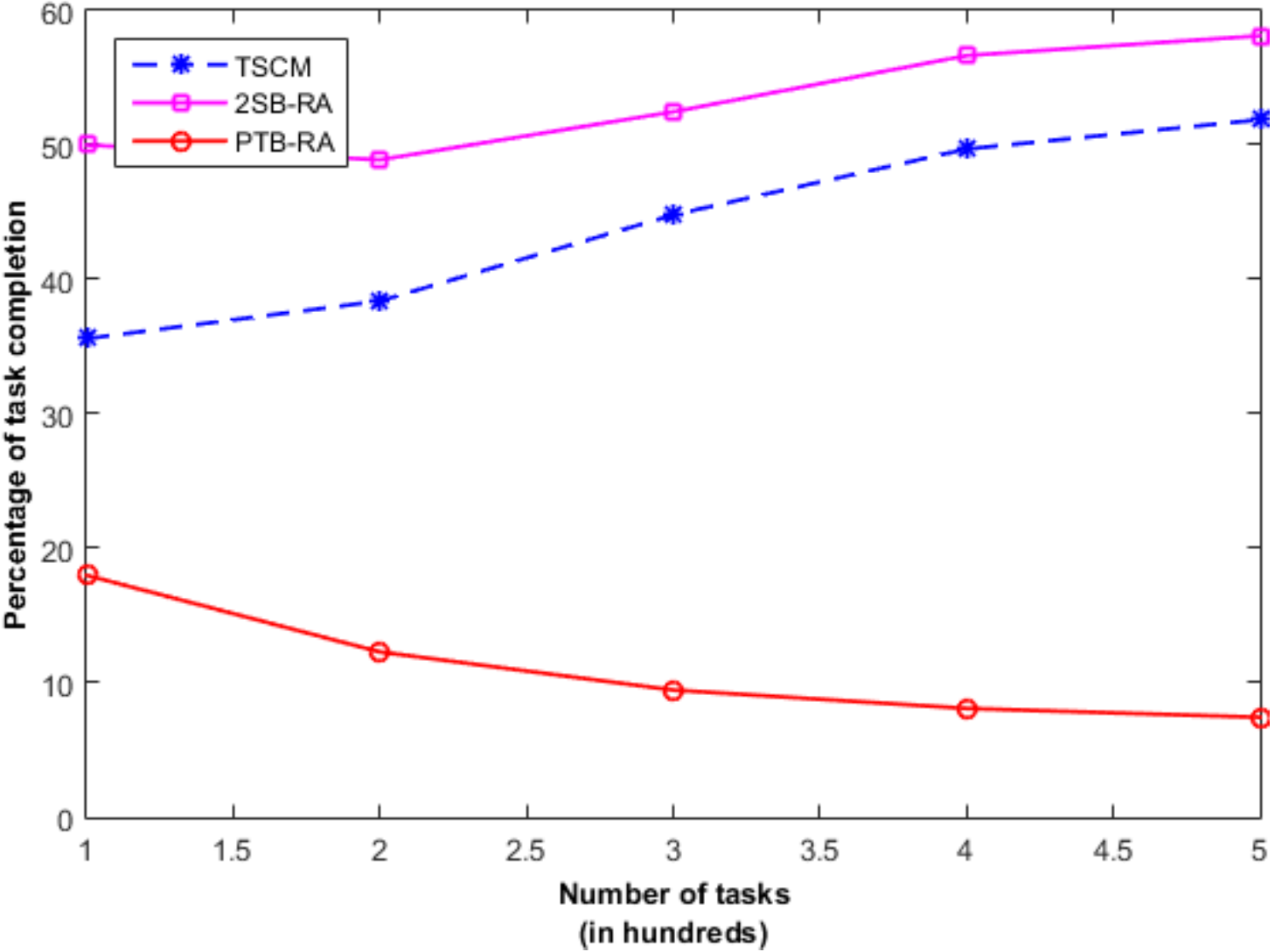}
		\caption{The impact of varying the number of tasks on the performance of PTB, $N=1000$.}
		\label{fig:ParamNumTaskPTB}
	\end{center}
\end{figure}

\begin{figure}[t]
	\begin{center}
		\setlength{\abovecaptionskip}{5pt plus 0pt minus 0pt}
		\setlength{\belowcaptionskip}{-10pt plus 0pt minus 0pt}
		\includegraphics[width=3.4in]{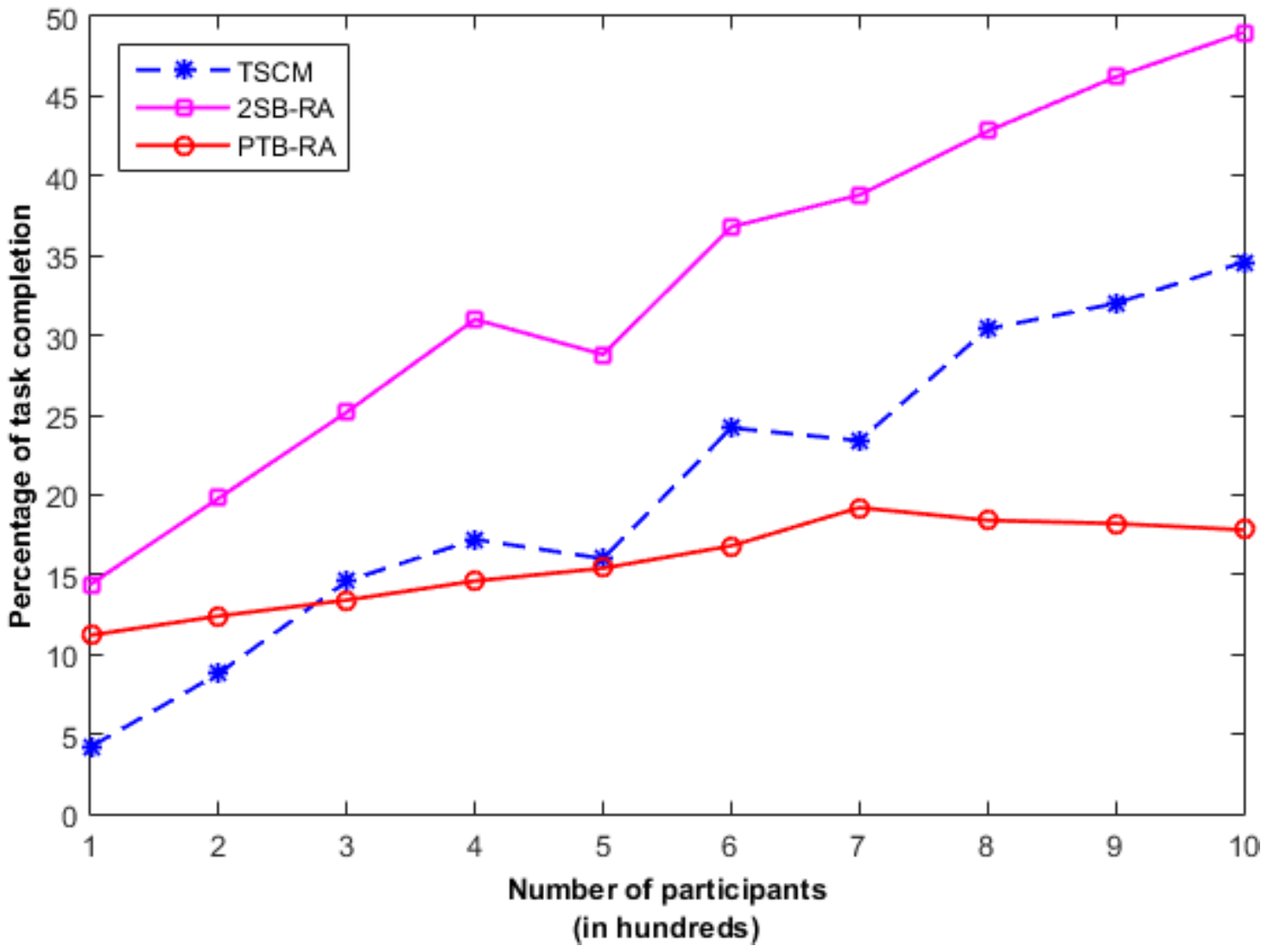}
		\caption{The impact of varying the number of participants on the performance of PTB, $M=100$.}
		\label{fig:ParamNumPartPTB}
	\end{center}
\end{figure}

\subsection{Comparing Per-task Bidding to two-stage Bidding}
To compare the performance of 2SB-RA and PTB-RA, we run ten-100 successive auctions with $N=100$ and $M=100$. We find that the PTB algorithm gives higher task completion ratios in an average of $1\%-2\%$ of the held auctions. This is depicted in Fig.~\ref{fig:Comp2SBPTB1}. 

\begin{figure}[t]
	\begin{center}
		\setlength{\abovecaptionskip}{5pt plus 0pt minus 0pt}
		\setlength{\belowcaptionskip}{-10pt plus 0pt minus 0pt}
		\includegraphics[width=3.4in]{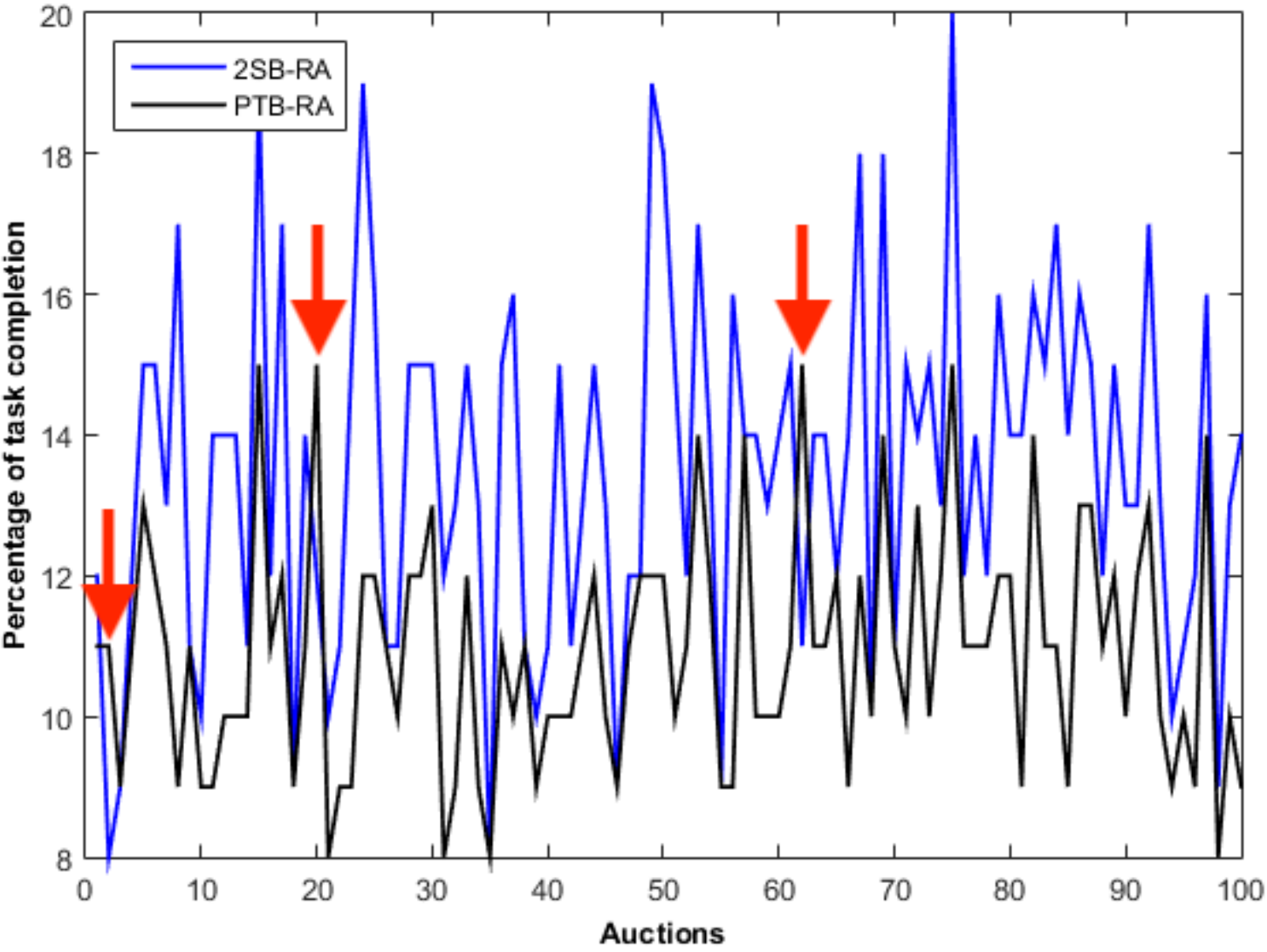}
		\caption{The performance of 2SB-RA and PTB-RA with regards to CR across $100$ successive auctions. The red arrows point to the locations where PTB outperformed 2SB.}
		\label{fig:Comp2SBPTB1}
	\end{center}
\end{figure}

For the reputation-unaware versions of the 2SB and the PTB algorithms, the performance of both of them has been found to be quite similar (with regards to the attained CR) when the number of tasks is set to $M\geq100$. When the number of tasks decreases at high values of $N$, the PTB outperforms the 2SB, and the gap between their performance increases proportionally with $N$. This is because the tasks in the former, are not allowed to be performed more than once, unlike the 2SB case. Through ten-100 successive auctions having 50 tasks and 300 participants, the PTB algorithm showed better performance in an average of $31.4\%$ of the held auctions. This percentage rises to $88.4\%$ when the number of participants increases to 500. In Fig.~\ref{fig:Comp2SBPTB3}, through $100$ successive auctions, we increased the number of participants to $N\geq600$; by then, the percentage rose to $100\%$ of the held auctions, i.e., all of the held auctions.

\noindent \textbf{Drawbacks of descriptive bidding-based auctions.} \hspace{1pt} It is worth mentioning that the proposed bidding procedures require considerably higher interaction from the participants, than previous techniques in the literature, due to the process of indicating a separate bid for every sensing task.


\begin{figure}[t]
	\begin{center}
		\setlength{\abovecaptionskip}{5pt plus 0pt minus 0pt}
		\setlength{\belowcaptionskip}{-10pt plus 0pt minus 0pt}
		\includegraphics[width=3.4in]{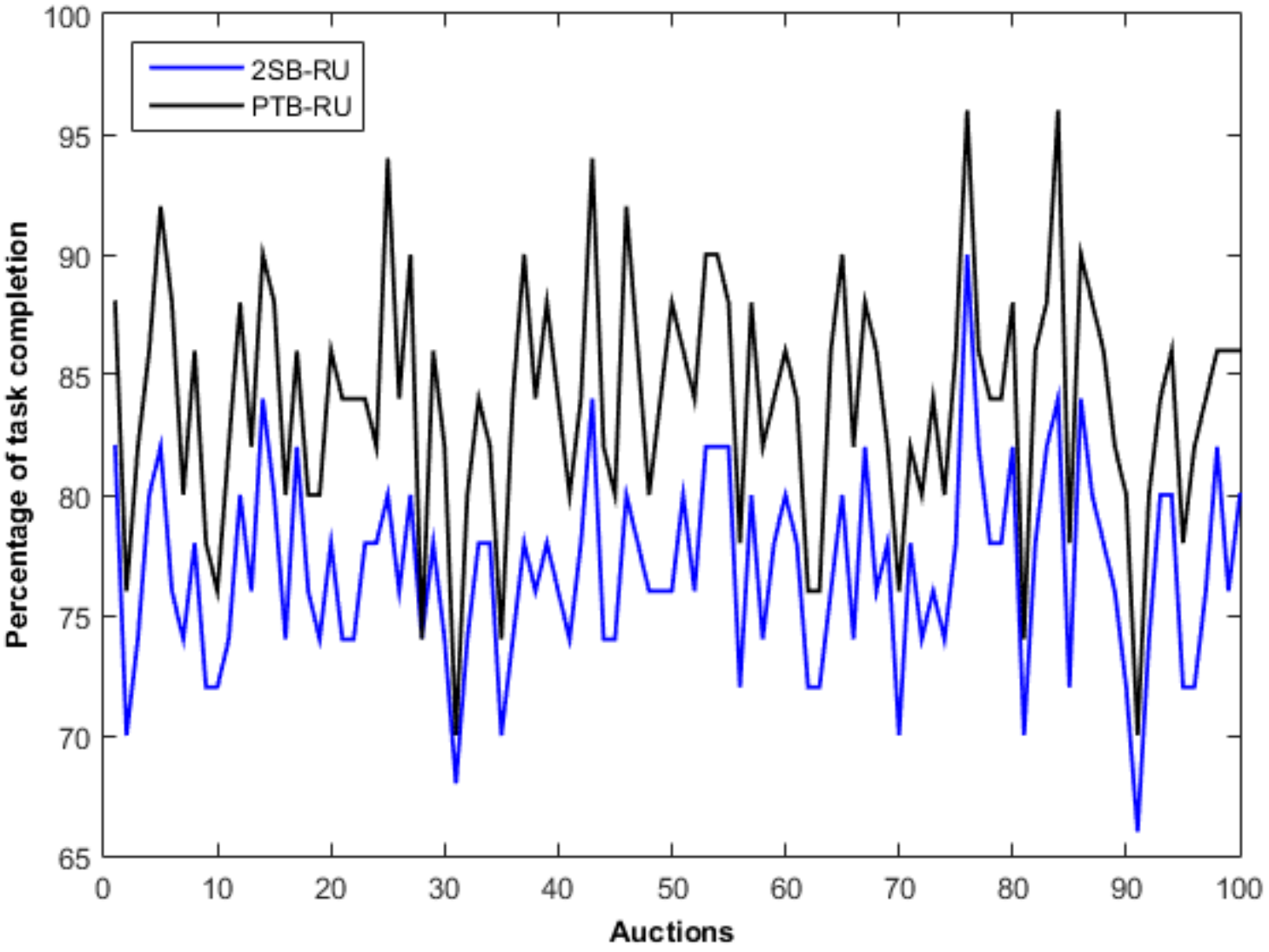}
		\caption{The performance of 2SB-RU and PTB-RU with regards to CR across $100$ successive auctions.}
		\label{fig:Comp2SBPTB3}
	\end{center}
\end{figure}

\section{Conclusion}
\label{sec:conclusion}
To the best of our knowledge, this research is the first to address the challenge of maximizing the CR in auction-based participatory MCS systems by proposing novel bidding procedures. The free parameters of these algorithms were identified and we simulated varying scenarios (varying number of auctions, tasks, and participants) in order to evaluate the effectiveness of the proposed techniques. Remarkable increase has been achieved, compared to the recent literature, with regards to the CR. Particularly, the 2SB algorithm has been demonstrated to consistently outperform the former reputation-aware techniques. Also, the user utility, under the proposed formulation, has been improved. Future work involves a new formulation for platform utility emphasizing the CR, in addition to learning models from synthetic data based on which the system can adaptively decide the bidding procedure that maximizes the CR over a window in time.

\end{document}